\documentclass[pra]{revtex4}
 \usepackage{amssymb} \usepackage{graphicx}

\begin{document}
 \title{Hidden symmetries and Lie algebra structures from geometric and supergravity Killing spinors}

\author{\"{O}zg\"{u}r A\c{c}{\i}k}
\email{ozacik@science.ankara.edu.tr, o.acik@lancaster.ac.uk}
\address{Department of Physics, Lancaster University,\\ Lancaster, LA1 4YB, United Kingdom\\}
\address{Department of Physics,
Ankara University,\\ Faculty of Sciences, 06100 Tando\u gan, Ankara,
Turkey}
\author{\"Umit Ertem}
 \email{umitertemm@gmail.com}
\address{School of Mathematics, The University of Edinburgh, James Clerk Maxwell Building, Peter Guthrie Tait Road, Edinburgh, EH9 3FD, United Kingdom\\}
\address{Department of Physics,
Ankara University,\\ Faculty of Sciences, 06100 Tando\u gan, Ankara,
Turkey}

\begin{abstract}
We consider geometric and supergravity Killing spinors and the spinor bilinears constructed out of them. The spinor bilinears of geometric Killing spinors correspond to the antisymmetric generalizations of Killing vector fields which are called Killing-Yano forms. They constitute a Lie superalgebra structure in constant curvature spacetimes. We show that the Dirac currents of geometric Killing spinors satisfy a Lie algebra structure up to a condition on 2-form spinor bilinears. We propose that the spinor bilinears of supergravity Killing spinors give way to different generalizations of Killing vector fields to higher degree forms. It is also shown that those supergravity Killing forms constitute a Lie algebra structure in six and ten dimensional cases. For five and eleven dimensional cases, the Lie algebra structure depends on an extra condition on supergravity Killing forms.
\end{abstract}

\maketitle

\section{Introduction}

Isometries of a manifold are generated by Killing vector fields. They correspond to the symmetries of the manifold and constitute a Lie algebra structure under the ordinary Lie bracket of vector fields. There are different types of generalizations of Killing vector fields to higher order tensors which are called hidden symmetries of the background. Symmetric tensor generalizations are called Killing tensors and they also satisfy a graded Lie algebra structure \cite{Dolan Kladouchou Card}. Killing tensors give rise to generalized constants of motion of the geodesic equation \cite{Crampin, Benn}. On the other hand, there are also antisymmetric tensor generalizations of Killing vectors which are called Killing-Yano (KY) forms \cite{Yano}. Although, in general they do not constitute a Lie algebra, in some special cases they satisfy a graded Lie algebra with respect to Schouten-Nijenhuis (SN) bracket \cite{Kastor Ray Traschen}. KY forms are also related to the constants of motion of the geodesic equation \cite{Hughston Penrose Sommers Walker}. Moreover, they are used in the separability properties of the Hamilton-Jacobi and Dirac equations \cite{Krtous Kubiznak Page Frolov, Cariglia Krtous Kubiznak, Cariglia Frolov Krtous Kubiznak, Cariglia} and construction of symmetry operators for the Dirac equation in curved backgrounds \cite{Benn Charlton, Benn Kress, Acik Ertem Onder Vercin}. G-structures which are used in the classification of supergravity backgrounds have also relations with KY forms \cite{Papadopoulos, Santillan}.

In a manifold with a spin structure, one can define spinor fields and a spinor covariant derivative acting on them. On some special manifolds, there can be special spinors that correspond to the solutions of some differential equations. One example is the geometric Killing spinors which are the solutions of the Killing spinor equation \cite{Lichnerowicz, Baum}. An important property of geometric Killing spinors is that one can obtain a Killing vector from the 1-form projection of the spinor bilinears constructed out of them. The metric dual of this 1-form projection is called the Dirac current of the spinor. Moreover, the higher degree parts of the spinor bilinears of geometric Killing spinors which are called $p$-form Dirac currents correspond to KY forms \cite{Acik Ertem}. Hence, the equations satisfied by the spinor bilinears of geometric Killing spinors define the equations of antisymmetric generalizations of Killing vector fields.

There are other ways of constructing Killing vectors from some special Killing spinors in supergravity theories. The bosonic supergravity backgrounds are Lorentzian spin manifolds admitting some special differential forms called fluxes. To obtain the bosonic supergravity backgrounds, the fermionic fields and their variations are taken to be zero \cite{Van Proeyen}. The vanishing conditions on the supersymmetry transformations of fermionic fields define some restriction equations for spinor parameters of the theory. These are called supergravity Killing spinor equations and they depend on the fluxes of the relevant supergravity theory. The variation with respect to gravitino field gives a differential condition on the spinor parameters and the variations with respect to other fermionic fields algebraically constrain them. The differential condition on the spinor parameters also give way to define a new spin connection on the background such that the spinor parameters are parallel with respect to this new connection. The 1-form components of the spinor bilinears of those supergravity Killing spinors are metrically related to the Killing vectors of the background. However, the higher degree projections of those spinor bilinears satisfy various equations which are generally different from the KY equation. They are related to G-structures and are used in the classification of supersymmetric supergravity backgrounds \cite{Gauntlett Gutowski Hull Pakis Reall, Gutowski Martelli Reall, Gauntlett Pakis}.

In this paper, we first consider the geometric Killing spinors and KY forms constructed out of them and show that besides the Lie superalgebra structure of KY forms, the Dirac currents of geometric Killing spinors can satisfy a Lie algebra under some conditions. Then, by considering supergravity Killing spinors, we propose that the equations satisfied by the spinor bilinears of supergravity Killing spinors can be considered as the equations of generalized Killing forms in supergravity backgrounds. Since they do not correspond to KY equation, they are somewhat different type of generalizations of Killing vectors. In both KY forms and supergravity Killing forms cases, the equations reduces to the Killing equation for 1-forms. So, although the 1-forms constructed out of both geometric Killing spinors and supergravity Killing spinors are metric duals of Killing vectors, they differ from each other in their higher degree sectors.  Furthermore, we show that those supergravity Killing forms have a Lie algebra structure in some cases. In particular, this arises in six-dimensional (1,0) supergravity, ten-dimensional type I, heterotic and type IIA supergravity Killing forms. For five dimensional and eleven dimensional supergravities, obtaining a new solution of the supergravity Killing form equation from the known two of them depend on an extra condition on the forms.

The paper is organized as follows. In section 2, we discuss the algebraic closure property of the set of KY forms in constant curvature spacetimes for the set of homogeneous forms ($p$-form Dirac currents) constructed out of geometric Killing spinors existing in the background. In a subsection, we show that the Dirac currents constructed out of geometric Killing spinors satisfy a Lie algebra structure up to a condition on their 2-form spinor bilinears. Some useful identities related to those Dirac currents are also given in the appendices. Section 3 focuses on the supergravity Killing spinors and their spinor bilinears. It is shown that the spinor bilinears of supergravity Killing spinors can be considered as different generalizations of Killing vector fields to higher degree forms and the supergravity Killing forms may constitute a Lie algebra in different cases. Section 4 concludes the paper.

\section{Geometric Killing Spinors and KY Forms}

Let us consider an $n$-dimensional manifold $M$ with a spin structure. This implies that one can define spinor fields as sections of the spinor bundle of $M$ and the spinor space will be denoted by $S$. Geometric Killing spinors are solutions of the following differential equation
\begin{equation}
\nabla_X\epsilon=\lambda\widetilde{X}.\epsilon
\end{equation}
where $\epsilon$ is a spinor, $X$ is a vector field, $\widetilde{X}$ is its metric dual and $\lambda$ is a real or pure imaginary number which is called the Killing number. $\nabla$ is the Levi-Civita spin connection and $.$ denotes the Clifford product. The existence of geometric Killing spinors on a spin manifold restricts the curvature characteristics of it in terms of the Killing number. For an $n$-dimensional spin manifold that carries a geometric Killing spinor, the scalar curvature is written as follows
\begin{equation}
{\cal{R}}=-4\lambda^2n(n-1)
\end{equation}
and the Clifford action of the curvature 2-forms $R_{ab}$ on the geometric Killing spinor is
\begin{equation}
R_{ab}.\epsilon=-4\lambda^2(e_a\wedge e_b).\epsilon
\end{equation}
where $e_a=\eta_{ab}e^b$ and $\{e^b\}$ is an orthonormal co-frame basis for the Lorentzian metric $\eta_{ab}e^a\otimes e^b$.

If we consider the spinor space $S$ and its dual $S^*$, then their tensor product $S\otimes S^*$  is the endomorphism algebra of $S$. Tensor product of a spinor $\epsilon$ and its dual $\bar{\epsilon}$ can be written as an inhomogeneous differential form
\begin{equation}
\epsilon\bar{\epsilon}=(\epsilon, \epsilon)+(\epsilon, e_a\epsilon)e^a+(\epsilon, e_{ba}\epsilon)e^{ab}+...+(\epsilon, e_{a_p...a_1}\epsilon)e^{a_1...a_p}+...+(-1)^{\lfloor n/2\rfloor}(\epsilon, z\epsilon)z
\end{equation}
where $( . , . )$ is the spinor inner product, $z$ is the volume form and $e^{a_1...a_p}:=e^{a_1}\wedge...\wedge e^{a_p}$. This equality is called the Fierz identity and the left hand side is the spinor bilinear constructed out of $\epsilon$ and the $p$-form components on the right hand side are called $p$-form Dirac currents or $p$-form components of the spinor bilinear. The 1-form spinor bilinear of a geometric Killing spinor is defined as
\begin{equation}
(\epsilon\bar{\epsilon})_1=(\epsilon, e_a\epsilon)e^a
\end{equation}
and the metric dual of it which is called the Dirac current is a Killing vector. The higher degree form spinor bilinears are defined similarly as
\begin{equation}
(\epsilon\bar{\epsilon})_p=(\epsilon, e_{a_p...a_1}\epsilon)e^{a_1...a_p}.
\end{equation}
Moreover, it is shown in \cite{Acik Ertem} that these $p$-form Dirac currents of geometric Killing spinors or their Hodge duals, depending on the data set determined by the Killing number, the inner product on spinors and the homogeneity degree $p$, satisfy the Killing-Yano equation
\begin{equation}
\nabla_X(\epsilon\bar{\epsilon})_p=\frac{1}{p+1}i_Xd(\epsilon\bar{\epsilon})_p
\end{equation}
where $i_X$ is the interior derivative with respect to $X$ and $d$ is the exterior derivative operator. This means that the $p$-form Dirac currents of geometric Killing spinors correspond to the generalizations of Killing vectors to totally antisymmetric tensors. This also shows that those spinor bilinears are a part of a Lie algebra structure in some special cases since the KY forms constitute a Lie superalgebra in constant curvature spacetimes. This curvature condition is consistent with the curvature condition that arises from the existence of geometric Killing spinors.

For a $p$-form $\omega_1$ and a $q$-form $\omega_2$ that are KY forms, their Schouten-Nijenhuis (SN) bracket which is defined as
\begin{equation}
[\omega_1,\omega_2]_{SN}=i_{X^a}\omega_1\wedge\nabla_{X_a}\omega_2+(-1)^{pq}i_{X^a}\omega_2\wedge\nabla_{X_a}\omega_1
\end{equation}
is also a KY form of degree $(p+q-1)$ in constant curvature spacetimes \cite{Kastor Ray Traschen}. This can be seen as follows. By taking the covariant derivative of the SN bracket of two KY forms $\omega_1$ and $\omega_2$, one obtains
\begin{eqnarray}
\nabla_{X_b}[\omega_1,\omega_2]_{SN}&=&\frac{1}{(p+1)(q+1)}\left(i_{X^a}i_{X_b}d\omega_1\wedge i_{X_a}d\omega_2+(-1)^{pq}i_{X^a}i_{X_b}d\omega_2\wedge i_{X_a}d\omega_1\right)\nonumber\\
&&+\frac{1}{q}i_{X^a}\omega_1\wedge i_{X_a}(R_{cb}\wedge i_{X^c}\omega_2)+\frac{(-1)^{pq}}{p}i_{X^a}\omega_2\wedge i_{X_a}(R_{cb}\wedge i_{X^c}\omega_1)
\end{eqnarray}
and applying the operator $i_{X_b}d$ to the SN bracket gives
\begin{eqnarray}
\frac{1}{p+q}i_{X_b}d[\omega_1,\omega_2]_{SN}&=&\frac{1}{(p+1)(q+1)}\left(i_{X^a}i_{X_b}d\omega_1\wedge i_{X_a}d\omega_2+(-1)^{pq}i_{X_a}i_{X_b}d\omega_2\wedge i_{X^a}d\omega_1\right)\nonumber\\
&&-\frac{(-1)^{p}}{p+q}\left(\frac{1}{p}+\frac{1}{q}\right)i_{X_b}\left(i_{X^a}\omega_1\wedge R_{ca}\wedge i_{X^c}\omega_2\right)
\end{eqnarray}
where we have used the following equality for KY $p$-forms
\begin{equation}
\nabla_{X_b}d\omega=\frac{p+1}{p}R_{ab}\wedge i_{X^a}\omega.
\end{equation}
In constant curvature spacetimes, we have $R_{ab}=ce_a\wedge e_b$ where $c$ is a constant. By comparing (9) and (10) in this case, one can see that the SN bracket of two KY forms also satisfy the KY equation;
\begin{equation}
\nabla_X[\omega_1,\omega_2]_{SN}=\frac{1}{p+q}i_Xd[\omega_1,\omega_2]_{SN}
\end{equation}
and in fact this bracket turns the space of KY forms into a Lie superalgebra. A Lie superalgebra $\mathfrak{g}=\mathfrak{g}_0\oplus\mathfrak{g}_1$ is a direct sum of a Lie algebra $\mathfrak{g}_0$ and a $\mathfrak{g}_0$-module $\mathfrak{g}_1$ with a (skew)-symmetric  bracket operation that satisfies the super-Jacobi identitiy. The $\mathfrak{g}_0$ and $\mathfrak{g}_1$ are called even and odd parts of the superalgebra respectively. A Lie bracket $[ . , . ]$ on a Lie superalgebra is defined as a bilinear multiplication
\begin{equation}
[ . , . ]:\mathfrak{g}_i\times\mathfrak{g}_j\longrightarrow\mathfrak{g}_{i+j}
\end{equation}
where $i, j=0, 1$ and satisfies the following skew-supersymmetry and super-Jacobi identities for the elements $a, b, c$ of $\mathfrak{g}$ and $|a|$ denotes the degree of $a$ which corresponds to 0 or 1 whether $a$ is in $\mathfrak{g}_0$ or $\mathfrak{g}_1$ respectively
\begin{eqnarray}
[a,b]&=&-(-1)^{|a||b|}[b,a]\nonumber\\
\left[a,[b,c]\right]&=&[[a, b], c]+(-1)^{|a||b|}[b, [a, c]].
\end{eqnarray}
By considering that definition let us denote the KY superalgebra as $\mathfrak{k}=\mathfrak{k}_0\oplus\mathfrak{k}_1$. The even part $\mathfrak{k}_0$ corresponds to the odd KY forms and the odd part $\mathfrak{k}_1$ corresponds to the even KY forms. That is because the SN bracket of two odd KY forms gives again an odd KY form:
\begin{equation}
[ . , . ]_{SN}:\mathfrak{k}_0\times\mathfrak{k}_0\longrightarrow\mathfrak{k}_0
\end{equation}
the SN bracket of an odd and an even KY forms gives and even KY form:
\begin{equation}
[ . , . ]_{SN}:\mathfrak{k}_0\times\mathfrak{k}_1\longrightarrow\mathfrak{k}_1
\end{equation}
and the SN bracket of two even KY forms gives and odd KY form:
\begin{equation}
[ . , . ]_{SN}:\mathfrak{k}_1\times\mathfrak{k}_1\longrightarrow\mathfrak{k}_0
\end{equation}
The SN bracket also satisfies the super-Jacobi identity. This extends the Lie algebra structure of Killing vector fields which is a subalgebra in $\mathfrak{k}_0$ to a Lie superalgebra structure of KY forms in constant curvature space-times.

\subsection{Lie Algebra Structure of Dirac Currents}

As a special case, let us consider the 1-form components of spinor bilinears of geometric Killing spinors. The metric duals of these 1-forms are called Dirac currents and correspond to the Killing vector fields. We denote the Dirac current of a spinor $\epsilon$ as $V_{\epsilon}=\widetilde{(\epsilon\bar{\epsilon})_1}$. Since Killing vector fields constitute a Lie algebra with respect to the Lie bracket of vector fields, one can ask that whether the Dirac currents of geometric Killing spinors satisfy a Lie subalgebra under this Lie bracket. From the definition of the Lie bracket of vector fields, one can write the Lie bracket of Dirac currents of geometric Killing spinors $\psi$ and $\phi$ as
\begin{eqnarray}
\widetilde{[V_{\psi}, V_{\phi}]}&=&{\cal{L}}_{V_{\psi}}\widetilde{V_{\phi}}
\end{eqnarray}
where ${\cal{L}}_X$ denotes the Lie derivative with respect to the vector field $X$ and it commutes with the metric duality operation when $X$ is a Killing vector field. The exterior and co-derivatives of different degree components of spinor bilinears of geometric Killing spinors can be written in terms of one higher or one lower degree components of spinor bilinears \cite{Acik Ertem}. If the Dirac current is a Killing vector, than all odd degree spinor bilinears are KY forms and satisfy the following equalities
\begin{eqnarray}
d(\psi\bar{\psi})_p&=&2\lambda(p+1)(\psi\bar{\psi})_{p+1}\nonumber\\
\delta(\psi\bar{\psi})_p&=&0.
\end{eqnarray}
In that case, the even degree spinor bilinears correspond to closed conformal KY forms and satisfy the following equalities
\begin{eqnarray}
d(\psi\bar{\psi})_p&=&0\nonumber\\
\delta(\psi\bar{\psi})_p&=&-2\lambda(n-p+1)(\psi\bar{\psi})_{p-1}
\end{eqnarray}
where $n$ is the dimension of spacetime. To obtain a Lie subalgebra of Dirac currents, the right hand side of (18) must be written as proportional to a Dirac current of another geometric Killing spinor $\zeta$;
\begin{equation}
\widetilde{[V_{\psi}, V_{\phi}]}=C_{\psi, \phi, \zeta}\widetilde{V_{\zeta}}
\end{equation}
where $C_{\psi, \phi, \zeta}$ are structure constants. Moreover, $\widetilde{V_{\zeta}}$ have to satisfy the both equalities in (19). Since the Lie bracket of two Killing vectors is always another Killing vector, it automatically satisfies the second equality;
\begin{equation}
\delta\widetilde{V_{\zeta}}=0,
\end{equation}
but this can also be deduced from the properties of the Dirac currents. On the other hand, by taking the exterior derivative of (21), one can see that
\begin{eqnarray}
C_{\psi, \phi, \zeta}d\widetilde{V_{\zeta}}&=&d{\cal{L}}_{V_{\psi}}\widetilde{V_{\phi}}\nonumber\\
&=&{\cal{L}}_{V_{\psi}}d\widetilde{V_{\phi}}\nonumber\\
&=&4\lambda{\cal{L}}_{V_{\psi}}(\phi\bar{\phi})_2
\end{eqnarray}
where we have used the commutativity of $d$ and ${\cal{L}}_X$ for any $X$ and the first equality of (19) for $\phi$. The Lie derivative of a differential form $\alpha$ with respect to a Killing vector field $K$ can be written in terms of the covariant derivative and the Clifford commutator as follows (see Appendix A)
\begin{equation}
{\cal{L}}_K\alpha=\nabla_K\alpha+\frac{1}{4}[d\widetilde{K}, \alpha]_{Cl}.
\end{equation}
Here $[ . , . ]_{Cl}$ is the Clifford bracket which is defined for arbitrary inhomogeneous differential forms $\alpha$ and $\beta$ as follows
\begin{equation}
[\alpha, \beta]_{Cl}=\alpha.\beta-\beta.\alpha.
\end{equation}
So, we can write (23) in the following form
\begin{equation}
C_{\psi, \phi, \zeta}d\widetilde{V_{\zeta}}=4\lambda\nabla_{V_{\psi}}(\phi\bar{\phi})_2+\lambda[d\widetilde{V_{\psi}}, (\phi\bar{\phi})_2]_{Cl}.
\end{equation}
We know that $(\phi\bar{\phi})_2$ is a closed conformal KY form which means that it satisfies the equation
\begin{equation}
\nabla_X*(\phi\bar{\phi})_2=\frac{1}{n-1}i_Xd*(\phi\bar{\phi})_2
\end{equation}
where * is the Hodge star operation and this can be written by using the metric compatibility of the Levi-Civita connection as
\begin{equation}
\nabla_X(\phi\bar{\phi})_2=\frac{1}{n-1}*^{-1}i_Xd*(\phi\bar{\phi})_2.
\end{equation}
Then, (26) turns into
\begin{eqnarray}
C_{\psi, \phi, \zeta}d\widetilde{V_{\zeta}}&=&\frac{4\lambda}{n-1}*^{-1}i_{V_{\psi}}d*(\phi\bar{\phi})_2+\lambda[d\widetilde{V_{\psi}}, (\phi\bar{\phi})_2]_{Cl}\nonumber\\
&=&-8\lambda^2*^{-1}i_{V_{\psi}}*\widetilde{V_{\phi}}+4\lambda^2[(\psi\bar{\psi})_2, (\phi\bar{\phi})_2]_{Cl}\nonumber\\
&=&-8\lambda^2\widetilde{V_{\phi}}\wedge\widetilde{V_{\psi}}+4\lambda^2[(\psi\bar{\psi})_2, (\phi\bar{\phi})_2]_{Cl}\nonumber\\
&=&4\lambda^2[\widetilde{V_{\psi}}, \widetilde{V_{\phi}}]_{Cl}+4\lambda^2[(\psi\bar{\psi})_2, (\phi\bar{\phi})_2]_{Cl}
\end{eqnarray}
where we have used (20) and (19) for $(\phi\bar{\phi})_2$ and $\widetilde{V_{\psi}}$ respectively. This shows that, the existence of a geometric Killing spinor $\zeta$ whose 2-form spinor-bilinear satisfies the following equality
\begin{equation}
(\zeta\bar{\zeta})_2=\frac{\lambda}{C_{\psi, \phi, \zeta}}\left([\widetilde{V_{\psi}}, \widetilde{V_{\phi}}]_{Cl}+[(\psi\bar{\psi})_2, (\phi\bar{\phi})_2]_{Cl}\right),
\end{equation}
with $V_{\zeta}=\frac{1}{C_{\psi, \phi, \zeta}}{[V_{\psi}, V_{\phi}]}$ its Dirac current is necessary and sufficient for closing the algebra restricted to the set of Dirac currents. Note that the closure of the set of Dirac currents with respect to the Lie bracket of vector fields is equivalent to the closure of the unified set of 1-form and 2-form Dirac currents with respect to the Clifford commutator apparent in (C6) and (30). A more aesthetic and conceptually satisfactory equality for the algebraic closure of this unified set is given by (C7). Namely, if the higher degree components of spinor bilinears of $\psi, \phi$ and $\zeta$ are related to each other as in (30), then the Dirac currents of them constitute a Lie algebra structure. Indeed, the Lie bracket of two Dirac currents is dependent on the choice of Killing numbers of two Killing spinors. It is shown in Appendix B that, to have a consistent Lie algebra structure, the Killing numbers of two spinors have to be equal. More relations and identities about Dirac currents are given in Appendix C.

\section{Supergravity Killing Forms and Lie Algebra Structures}

Supergravity theories are generalizations of General Relativity in various dimensions to incorporate the supersymmetry transformations \cite{Freedman Van Proeyen}. Besides the bosonic gravitational degrees of freedom, they also include fermionic matter degrees of freedom and some extra bosonic fields to compansate the fermionic degrees of freedom. Fermionic and bosonic fields of the theory transform to each other under supersymmetry transformations. For a spinor parameter $\epsilon$, the bosonic field $\phi$ transforms under supersymmetry transformations to a fermionic field $\psi$ and vice versa as follows
\begin{eqnarray}
\delta(\epsilon)\phi&=&\bar{\epsilon}\psi\nonumber\\
\delta(\epsilon)\psi&=&(\nabla+f(\phi))\epsilon.
\end{eqnarray}
where $\delta$ denotes the variation. Bosonic sectors of supergravity theories are obtained by taking the fermionic fields to be zero. In that case, variations of the action with respect to the fermionic fields give some constraints on the spinor parameters that are used in the supersymmetry transformations. For example, the variation of the gravitino gives different differential constraints on the spinor parameter in different supergravity theories and the constraints also include the extra bosonic fields. On the other hand, variations with respect to the other fermionic fields give some algebraic conditions on the spinor parameter. These constraints are called supergravity Killing spinor equations. We consider the differential Killing spinor equations in various dimensions and discuss the properties of the solutions of the equations that are satisfied by the spinor bilinears constructed out of supergravity Killing spinors. These supergravity Killing form equations are then different generalizations of the Killing equation to higher degree forms.

\subsection{Six and Ten Dimensional Supergravities}

We first consider the six dimensional case. In six dimensions with Lorenzian signature, the Clifford algebra corresponds to $Cl(5,1)\cong\mathbb{H}(4)$ where $\mathbb{H}(4)$ denotes the $4\times 4$ matrices with quaternionic entries. Spinors are elements of the irreducible representations of the spin group $Spin(5,1)$ and in six dimensions there are two irreducible representations whose elements are called chiral spinors. In six-dimensional $(1,0)$ supergravity, the spinor parameter $\epsilon$ is a chiral spinor and the Killing spinor equation is given by
\begin{equation}
D_X\epsilon=\nabla_X\epsilon+\frac{1}{4}i_XH.\epsilon=0
\end{equation}
where $H$ is an anti-self-dual closed 3-form and $X$ is an arbitrary vector field. $D$ is the supergravity spin connection defined in terms of the Levi-Civita spin connection $\nabla$ and the bosonic field $H$. This supergravity spin connection can be induced from a Clifford bundle connection which is flat and with non-zero torsion that corresponds to the field $H$ \cite{Gutowski Martelli Reall, Chamseddine OFarrill Sabra, OFarrill Hustler}.

So, the action of the spin connection on a supergravity Killing spinor is written as follows
\begin{equation}
\nabla_X\epsilon=-\frac{1}{4}i_XH.\epsilon
\end{equation}
The equation satisfied by the spinor bilinears of supergravity Killing spinors in six dimensions is determined by the supergravity Killing spinor equation. Since the Levi-Civita connection respects the degree of a form and is compatible with the spinor duality operation, one can write
\begin{eqnarray}
\nabla_X(\epsilon\bar{\epsilon})_p&=&\left((\nabla_X\epsilon)\bar{\epsilon}\right)_p+\left(\epsilon(\nabla_X\bar{\epsilon})\right)_p\\
&=&\left((\nabla_X\epsilon)\bar{\epsilon}\right)_p+\left(\epsilon(\overline{\nabla_X\epsilon})\right)_p.
\end{eqnarray}
By using (33), we obtain
\begin{eqnarray}
\nabla_X(\epsilon\bar{\epsilon})_p&=&-\frac{1}{4}\left(i_XH.\epsilon\bar{\epsilon}\right)_p-\frac{1}{4}\left(\epsilon\overline{i_XH.\epsilon}\right)_p\\
&=&-\frac{1}{4}\left(i_XH.\epsilon\bar{\epsilon}\right)_p+\frac{1}{4}\left(\epsilon\bar{\epsilon}.i_XH\right)_p
\end{eqnarray}
where we have used $\overline{i_XH.\epsilon}=-\bar{\epsilon}.i_XH$ since $i_XH$ is  a 2-form and a 2-form has always got a minus sign under the involutions of the Clifford algebra. Hence, the equation satisfied by the spinor bilinears of Killing spinors can be written in the following form
\begin{eqnarray}
\nabla_X(\epsilon\bar{\epsilon})_p&=&-\frac{1}{4}\left([i_XH, \epsilon\bar{\epsilon}]_{Cl}\right)_p\nonumber\\
&=&-\frac{1}{4}\left[i_XH, (\epsilon\bar{\epsilon})_p\right]_{Cl}
\end{eqnarray}
by using the degree preserving property of the Clifford commutator of a 2-form with an arbitrary form. On the other hand, the Clifford commutator can be written in terms of wedge products as
\begin{eqnarray}
[\alpha,\beta]_{Cl}=\sum_{k=0}^{n}\frac{(-1)^{\lfloor{k/2\rfloor}}}{k!}\Bigg\{\left(\eta^k i_{X_{I(k)}}\alpha\right)\wedge i_{X^{I(k)}}\beta-\left(\eta^k i_{X_{I(k)}}\beta\right)\wedge i_{X^{I(k)}}\alpha\Bigg\}
\end{eqnarray}
where $\lfloor{\rfloor}$ is the floor function that takes the integer part of the argument, $\eta$ is the main automorphism of the exterior algebra that acts on a $p$-form $\alpha$ as $\eta\alpha=(-1)^p\alpha$ and $I(k)$ is a multi index (see \cite{Benn Tucker, Tucker} and Appendix A of \cite{Acik Ertem}). In (38), $i_XH$ is a 2-form and the Clifford bracket on the right hand side of (38) reduces to the following equality
\begin{equation}
\left([i_XH, \epsilon\bar{\epsilon}]_{Cl}\right)_p=-2i_{X_a}i_XH\wedge i_{X^a}(\epsilon\bar{\epsilon})_p.
\end{equation}
If we write (38) for $(\epsilon\bar{\epsilon})_1$ we find
\begin{eqnarray}
\nabla_{X_b}(\epsilon\bar{\epsilon})_1&=&\frac{1}{2}i_{X_a}i_{X_b}H\wedge i_{X^a}(\epsilon\bar{\epsilon})_1
\end{eqnarray}
and wedging (41) with $e^b$ gives
\begin{equation}
d(\epsilon\bar{\epsilon})_1=-i_{X_a}H\wedge i_{X^a}(\epsilon\bar{\epsilon})_1.
\end{equation}
By comparing (41) and (42), one sees that
\begin{equation}
\nabla_X(\epsilon\bar{\epsilon})_1=\frac{1}{2}i_Xd(\epsilon\bar{\epsilon})_1
\end{equation}
which is the Killing equation as can be expected since it is known that the Dirac current of a supergravity Killing spinor is a Killing vector. This is not true for higher degree bilinear forms, namely they do not satisfy the KY equation. For example, it can be seen from (38) that the other non-zero spinor bilinear $(\epsilon\bar{\epsilon})_3$ satisfies the following equalities
\begin{eqnarray}
\nabla_{X_b}(\epsilon\bar{\epsilon})_3&=&\frac{1}{2}i_{X_a}i_{X_b}H\wedge i_{X^a}(\epsilon\bar{\epsilon})_3\\
d(\epsilon\bar{\epsilon})_3&=&-i_{X_a}H\wedge i_{X^a}(\epsilon\bar{\epsilon})_3
\end{eqnarray}
and by comparing them, one obtains
\begin{eqnarray}
\nabla_{X_b}(\epsilon\bar{\epsilon})_3&=&\frac{1}{2}\left[i_{X_b}d(\epsilon\bar{\epsilon})_3+i_{X_a}H\wedge i_{X_b}i_{X^a}(\epsilon\bar{\epsilon})_3\right]\nonumber\\
&=&\frac{1}{2}i_{X_b}d(\epsilon\bar{\epsilon})_3+\frac{1}{4}[H, i_{X_b}(\epsilon\bar{\epsilon})_3]_{Cl}.
\end{eqnarray}
However, this means that (38) is another generalization of the Killing equation to higher degree forms in six dimensional supergravity backgrounds, since it reduces to the metric duals of Killing vector fields in the 1-form case. So, it can be called as the supergravity Killing form equation.  Moreover, the solutions of (38) constitute a Lie algebra structure. That equation can be written for a general $p$-form $\omega$ as follows
\begin{equation}
\nabla_X\omega=-\frac{1}{4}[i_XH, \omega]_{Cl}.
\end{equation}
Let $\omega_1$ and $\omega_2$ be solutions of (47), then their Clifford bracket which is the Clifford commutator of those forms is also a solution. This can be seen in the following way
\begin{eqnarray}
\nabla_X[\omega_1, \omega_2]_{Cl}&=&[\nabla_X\omega_1,\omega_2]_{Cl}+[\omega_1, \nabla_X\omega_2]_{Cl}\nonumber\\
&=&-\frac{1}{4}[[i_XH, \omega_1]_{Cl},\omega_2]_{Cl}-\frac{1}{4}[\omega_1, [i_XH, \omega_2]_{Cl}]_{Cl}\nonumber\\
&=&-\frac{1}{4}[i_XH.\omega_1, \omega_2]_{Cl}+\frac{1}{4}[\omega_1.i_XH, \omega_2]_{Cl}\nonumber\\
&&-\frac{1}{4}[\omega_1, i_XH.\omega_2]_{Cl}+\frac{1}{4}[\omega_1, \omega_2.i_XH]_{Cl}\nonumber\\
&=&-\frac{1}{4}\left(i_XH.\omega_1.\omega_2+\omega_2.\omega_1.i_XH-i_XH.\omega_2.\omega_1-\omega_1.\omega_2.i_XH\right)\nonumber\\
&=&-\frac{1}{4}[i_XH, [\omega_1, \omega_2]_{Cl}]_{Cl}.
\end{eqnarray}
Hence, solutions of (47) have a Lie algebra structure with Clifford bracket $[ . , . ]_{Cl}$ since this bracket satisfy the following Lie bracket properties
\begin{equation}
[\omega_1, \omega_2]_{Cl}=-[\omega_2, \omega_1]_{Cl}
\end{equation}
\begin{equation}
[\omega_1, [\omega_2, \omega_3]_{Cl}]_{Cl}+[\omega_2, [\omega_3, \omega_1]_{Cl}]_{Cl}+[\omega_3, [\omega_1, \omega_2]_{Cl}]_{Cl}=0.
\end{equation}

The similar construction of supergravity Killing forms can also be obtained in ten dimensional supergravity backgrounds. In ten dimensional heterotic supergravity, the differential Killing spinor equation is exactly the same as (32). The spinor parameter $\epsilon$ is a chiral spinor and $H$ is again a closed 3-form which is not anti-self dual now \cite{Gran Lohrmann Papadopoulos, OFarrill HackettJones Moutsopoulos}. Moreover, the spinor parameter satisfies some additional algebraic constraints which are given by
\begin{eqnarray}
\left(d\phi+\frac{1}{2}H\right).\epsilon&=&0\\
F_2.\epsilon&=&0
\end{eqnarray}
where $\phi$ is the dilaton scalar field and $F_2$ is the 2-form field. However, these algebraic constraints do not affect the construction of Lie algebra structures of supergravity Killing forms since they do not appear in the calculations. So, the six dimensional construction can exactly be applied to the heterotic theory and supergravity Killing forms of heterotic backgrounds with a Lie algebra structure can again be found as generalizations of Killing vector fields. This also works for the differential Killing spinor equation of ten dimensional type I supergravity.
On the other hand, for type IIA supergravity the differential Killing spinor equation is the following
\begin{equation}
\nabla_X\epsilon=-\frac{1}{4}i_XH.\epsilon-\frac{1}{8}e^{\phi}i_XF_2.\epsilon+\frac{1}{8}e^{\phi}(\widetilde{X}\wedge F_4).\epsilon
\end{equation}
again $\phi$ is the dilaton and $F_2$, $F_4$ are Ramond-Ramond (RR) fields \cite{Duff Nilsson Pope}. When RR fields vanish, the equation reduces to the six dimensional and ten dimensional heterotic theories and the construction of supergravity Killing forms as generalizations of Killing vector fields in those backgrounds can be done by the same procedure. For non-zero RR fields, the covariant derivative of the spinor bilinears constructed out of the solutions of (53) will satisfy
\begin{eqnarray}
\nabla_X\left(\epsilon\bar{\epsilon}\right)_p&=&\left((\nabla_X\epsilon)\bar{\epsilon}\right)_p+\left(\epsilon(\nabla_X\bar{\epsilon})\right)_p\nonumber\\
&=&-\frac{1}{4}\left(i_XH.\epsilon\bar{\epsilon}\right)_p-\frac{e^{\phi}}{8}\left(i_XF_2.\epsilon\bar{\epsilon}\right)_p+\frac{e^{\phi}}{8}\left((\widetilde{X}\wedge F_4).\epsilon\bar{\epsilon}\right)_p\nonumber\\
&&-\frac{1}{4}\left(\epsilon\overline{i_XH.\epsilon}\right)_p-\frac{e^{\phi}}{8}\left(\epsilon\overline{i_XF_2.\epsilon}\right)_p+\frac{e^{\phi}}{8}\left(\epsilon\overline{(\widetilde{X}\wedge F_4).\epsilon}\right)_p.
\end{eqnarray}
The duality operation $\bar{\quad}$ on Clifford algebra elements can be defined by using the involutions of the algebra. So, we have
\begin{eqnarray}
\overline{i_XH.\epsilon}&=&\bar{\epsilon}.\left(i_XH\right)^{\xi\eta}=-\bar{\epsilon}.i_XH\nonumber\\
\overline{i_XF_2.\epsilon}&=&\bar{\epsilon}.\left(i_XF_2\right)^{\xi\eta}=-\bar{\epsilon}.i_XF_2\nonumber\\
\overline{(\widetilde{X}\wedge F_4).\epsilon}&=&\bar{\epsilon}.\left(\widetilde{X}\wedge F_4\right)^{\xi\eta}=-\bar{\epsilon}.\left(\widetilde{X}\wedge F_4\right)\nonumber
\end{eqnarray}
where for a $p$-form $\alpha$, the involution $\xi$ acts as $\alpha^{\xi}=(-1)^{\lfloor p/2\rfloor}\alpha$. Then, (54) transforms into
\begin{equation}
\nabla_X\left(\epsilon\bar{\epsilon}\right)_p=-\frac{1}{8}\left(\left[2i_XH+e^{\phi}i_XF_2-e^{\phi}\widetilde{X}\wedge F_4,\epsilon\bar{\epsilon}\right]_{Cl}\right)_p.
\end{equation}
Although the Clifford commutator of a 2-form with an arbitrary form preserves the degree of the form, this is not true for the Clifford commutators of a 1-form and 5-form with an arbitrary form. So, we cannot use the trick in (38) to obtain a homogeneous degree equation. Instead, since the Dirac currents of supergravity Killing spinors correspond to Killing vectors, we have the following generalization of the Killing equation as the supergravity Killing form equation for an inhomogeneous form $\omega$
\begin{equation}
\nabla_X\omega=-\frac{1}{8}\left(\left[2i_XH+e^{\phi}i_XF_2-e^{\phi}\widetilde{X}\wedge F_4,\omega\right]_{Cl}\right).
\end{equation}
By denoting $\Xi_X:=2i_XH+e^{\phi}i_XF_2-e^{\phi}\widetilde{X}\wedge F_4$, we can check the Lie algebra structure of the solutions of (56). For two solutions $\omega_1$ and $\omega_2$, the covariant derivative of their Clifford bracket gives
\begin{eqnarray}
\nabla_X[\omega_1,\omega_2]_{Cl}&=&[\nabla_X\omega_1,\omega_2]_{Cl}+[\omega_1,\nabla_X\omega_2]_{Cl}\nonumber\\
&=&-\frac{1}{8}\left[[\Xi_X,\omega_1]_{Cl},\omega_2\right]_{Cl}-\frac{1}{8}\left[\omega_1,[\Xi_X,\omega_2]_{Cl}\right]_{Cl}\nonumber\\
&=&-\frac{1}{8}\left[\Xi_X,[\omega_1,\omega_2]_{Cl}\right]_{Cl}
\end{eqnarray}
Hence, the solutions of (56) also have a Lie algebra structure with the Clifford bracket.

\subsection{Five and Eleven Dimensional Supergravities}

In five dimensions with Lorenzian signature, the Clifford algebra corresponds to $Cl(4,1)\cong\mathbb{C}(4)$ where $\mathbb{C}(4)$ denotes the $4\times 4$ matrices with complex entries. Spinors are elements of the irreducible representations of the spin group $Spin(4,1)$ and in five dimensions there is a unique irreducible representation whose elements are Dirac spinors. The differential Killing spinor equation in five-dimensional supergravity takes the following form
\begin{equation}
D_X\epsilon=\nabla_X\epsilon+\frac{1}{\sqrt{3}}\left(i_XF-\frac{1}{2}\widetilde{X}\wedge F\right).\epsilon=0
\end{equation}
where $F$ is a closed 2-form and $X$ is an arbitrary vector field \cite{Gauntlett Gutowski Hull Pakis Reall}. The supergravity spin connection $D$ is defined in terms of the Levi-Civita spin connection $\nabla$ and the 2-form field $F$. However, in that case the supergravity spin connection can not be induced from a Clifford bundle connection. So, the action of the spin connection induced by the Levi-Civita connection on the spinor parameter $\epsilon$ is written as
\begin{equation}
\nabla_X\epsilon=-\frac{1}{\sqrt{3}}\left(i_XF-\frac{1}{2}\widetilde{X}\wedge F\right).\epsilon
\end{equation}
By using the following decompositions of the Clifford product of a 1-form with an arbitrary form in terms of wedge product and interior derivative
\begin{eqnarray}
\widetilde{X}.F&=&\widetilde{X}\wedge F+i_XF\\
F.\widetilde{X}&=&\widetilde{X}\wedge F-i_XF
\end{eqnarray}
one can write the Killing spinor equation as follows
\begin{equation}
\nabla_X\epsilon=-\frac{1}{4\sqrt{3}}\left(\widetilde{X}.F-3F.\widetilde{X}\right).\epsilon
\end{equation}
So, the covariant derivative of spinor bilinears constructed out of supergravity Killing spinors can be found as
\begin{eqnarray}
\nabla_X(\epsilon\bar{\epsilon})_p&=&\left((\nabla_X\epsilon)\bar{\epsilon}\right)_p+\left(\epsilon(\overline{\nabla_X\epsilon})\right)_p\nonumber\\
&=&-\frac{1}{4\sqrt{3}}\left(\left(\widetilde{X}.F-3F.\widetilde{X}\right).\epsilon\bar{\epsilon}\right)_p-\frac{1}{4\sqrt{3}}\left(\epsilon\overline{\left(\widetilde{X}.F-3F.\widetilde{X}\right).\epsilon}\right)_p\nonumber\\
&=&-\frac{1}{4\sqrt{3}}\left(\left(\widetilde{X}.F-3F.\widetilde{X}\right).\epsilon\bar{\epsilon}+\epsilon\bar{\epsilon}.\left(F.\widetilde{X}-3\widetilde{X}.F\right)\right)_p\nonumber\\
&=&-\frac{1}{4\sqrt{3}}\left(\left[\left(\widetilde{X}.F-3F.\widetilde{X}\right), \epsilon\bar{\epsilon}\right]_{Cl}\right)_p+\frac{1}{2\sqrt{3}}\left(\epsilon\bar{\epsilon}.(F.\widetilde{X}+\widetilde{X}.F)\right)_p
\end{eqnarray}
here we have used that $\overline{\left(\widetilde{X}.F-3F.\widetilde{X}\right).\epsilon}=\bar{\epsilon}.\left(\widetilde{X}.F-3F.\widetilde{X}\right)^{\xi\eta}=\bar{\epsilon}.\left(F.\widetilde{X}-3\widetilde{X}.F\right)$ since for the Clifford product of two Clifford forms $\alpha$ and $\beta$, we have $(\alpha.\beta)^{\xi\eta}=\beta^{\xi\eta}.\alpha^{\xi\eta}$.

The Dirac currents of supergravity Killing spinors in five dimensions are Killing vector fields \cite{Gauntlett Gutowski Hull Pakis Reall}. So, the 1-form part of (63) corresponds to the Killing equation. Thus, it can be seen as a generalization of Killing vector fields to higher degree objects in the supergravity background and the general supergravity Killing form equation for an inhomogeneous form $\omega$ can be written in the following form
\begin{equation}
\nabla_X\omega=-\frac{1}{4\sqrt{3}}\left[\left(\widetilde{X}.F-3F.\widetilde{X}\right), \omega\right]_{Cl}+\frac{1}{2\sqrt{3}}\omega.(F.\widetilde{X}+\widetilde{X}.F).
\end{equation}
Indeed, the second term on the right hand side reflects the fact that the supergravity spin connection $D$ can not be induced from a Clifford bundle connection. If that were the case, we would have only the first term on the right hand side which is a Clifford commutator as in the six dimensional case (47). By considering the 3-form $*F$ as torsion, another type of generalization of Killing vectors to higher degree forms can also be obtained \cite{Kubiznak Kunduri Yasui}. This generalized KY equation corresponds to the KY equation with a connection with torsion and is an equation for homogeneous degree forms. Moreover, it is not related to the spinor bilinears of supergravity Killing spinors. However, the generalization that we define in (64) is an inhomogeneous equation and constructed out of spinor bilinears. So, it is a different generalization of the Killing equation and in general it does not correspond to the generalized KY equation defined in \cite{Kubiznak Kunduri Yasui}.

To search for a Lie algebra structure of the solutions of (64), let us define the following equalities 
\begin{eqnarray}
\Phi_X&:=&\widetilde{X}.F-3F.\widetilde{X}\\
\Psi_X&:=&F.\widetilde{X}+\widetilde{X}.F
\end{eqnarray}
and consider the Clifford bracket of two solutions $\omega_1$ and $\omega_2$ to compute the covariant derivative
\begin{eqnarray}
\nabla_X\left[\omega_1, \omega_2\right]_{Cl}&=&\left[\nabla_X\omega_1, \omega_2\right]_{Cl}+\left[\omega_1, \nabla_X\omega_2\right]_{Cl}\nonumber\\
&=&-\frac{1}{4\sqrt{3}}\left[\left[\Phi_X, \omega_1\right]_{Cl}, \omega_2\right]_{Cl}+\frac{1}{2\sqrt{3}}\left[\omega_1.\Psi_X, \omega_2\right]_{Cl}\nonumber\\
&&-\frac{1}{4\sqrt{3}}\left[\omega_1, \left[\Phi_X, \omega_2\right]_{Cl}\right]_{Cl}+\frac{1}{2\sqrt{3}}\left[\omega_1, \omega_2.\Psi_X\right]_{Cl}\nonumber\\
&=&-\frac{1}{4\sqrt{3}}\left[\Phi_X, \left[\omega_1, \omega_2\right]_{Cl}\right]_{Cl}+\frac{1}{2\sqrt{3}}\left[\omega_1, \omega_2\right]_{Cl}.\Psi_X+\frac{1}{2\sqrt{3}}\left(\omega_1.\Psi_X.\omega_2-\omega_2.\Psi_X.\omega_1\right).
\end{eqnarray}
The first two terms on the right hand side are expected from the supergravity Killing form equation. Because of the third term on the right hand side, there is no general Lie algebra structure for the solutions. But, for some $\omega_1$ and $\omega_2$ that satisfy the following condition
\begin{equation}
\omega_1.\Psi_X.\omega_2=\omega_2.\Psi_X.\omega_1
\end{equation}
their Clifford bracket is also a solution for the supergravity Killing form equation.

The same construction is also possible for eleven dimensional supergravity. There, the differential Killing spinor equation only differs by a constant factor from (59) \cite{Cremmer Julia Scherk, OFarrill Papadopoulos}
\begin{equation}
\nabla_X\epsilon=-\frac{1}{6}\left(i_XF-\frac{1}{2}\widetilde{X}\wedge F\right).\epsilon
\end{equation}
where $F$ is a 4-form here. This equation can be written in terms of Clifford products as
\begin{equation}
\nabla_X\epsilon=-\frac{1}{24}\left(\widetilde{X}.F-3F.\widetilde{X}\right).\epsilon
\end{equation}
Similar procedures can be applied to obtain the covariant derivative of spinor bilinears as in five dimensional case. However, the duality operation gives the following result in this case $\overline{\left(\widetilde{X}.F-3F.\widetilde{X}\right).\epsilon}=-\bar{\epsilon}.\left(F.\widetilde{X}-3\widetilde{X}.F\right)$ since $F$ is a 4-form. So, we have
\begin{equation}
\nabla_X(\epsilon\bar{\epsilon})_p=-\frac{1}{24}\left(\left[\left(\widetilde{X}.F-3F.\widetilde{X}\right), \epsilon\bar{\epsilon}\right]_{Cl}\right)_p-\frac{1}{6}\left(\epsilon\bar{\epsilon}.[F,\widetilde{X}]_{Cl}\right)_p
\end{equation}
and supergravity Killing form equation for an inhomogeneous $\omega$ is written as
\begin{equation}
\nabla_X\omega=-\frac{1}{24}\left[\left(\widetilde{X}.F-3F.\widetilde{X}\right),\omega\right]_{Cl}-\frac{1}{6}\omega.[F,\widetilde{X}]_{Cl}.
\end{equation}
By defining
\begin{equation}
\Psi'_X:=[F,\widetilde{X}]_{Cl}
\end{equation}
we obtain
\begin{equation}
\nabla_X\left[\omega_1, \omega_2\right]_{Cl}=-\frac{1}{24}\left[\Phi_X, \left[\omega_1, \omega_2\right]_{Cl}\right]_{Cl}-\frac{1}{6}\left[\omega_1, \omega_2\right]_{Cl}.\Psi'_X-\frac{1}{6}\left(\omega_1.\Psi'_X.\omega_2-\omega_2.\Psi'_X.\omega_1\right)
\end{equation}
and the solutions of (72) have a Lie algebra structure under $[,]_{Cl}$ if the following condition is satisfied
\begin{equation}
\omega_1.\Psi'_X.\omega_2=\omega_2.\Psi'_X.\omega_1
\end{equation}
Note that in both five and eleven dimensional cases, the extra conditions on having a Lie algebra structure is written in terms of a 3-form. In five dimensions, we have $\Psi_X=2\widetilde{X}\wedge F$ and in eleven dimensions we have $\Psi'_X=-2i_XF$ which are both 3-forms. This means that (68) and (75) are non-trivial conditions on $\omega_1$ and $\omega_2$.

\section{Conclusion}

On a spin manifold with a special structure, the generalizations of Killing vector fields to higher degree forms can be constructed through some spinors that are called Killing spinors. Spinor bilinears of those special spinors satisfy a differential equation and this gives the definitive equation for the higher degree generalizations of Killing vector fields. Spinor bilinears of geometric Killing spinors correspond to KY forms and they built up a Lie superalgebra in constant curvature spacetimes. Moreover, one can find a Lie algebra structure of the Dirac currents for geometric Killing spinors up to a condition on the 2-form spinor bilinears of those spinors. In the analysis made for the algebraic closure of the set of Dirac currents of geometric Killing spinors, it is reasonable to ask whether it is possible to write an exact equality for the third Killing spinor in terms of the former two. Even for the existence of enough number of Killing spinors, answering the above question is still not trivial and requires further investigation. The appearance of and the role played by the spinorial Lie derivatives in constructing the conditions for the closure property for the set of Killing spinors put some restrictions that have to be overcomed. That is because, it seems hard to find a symmetric product for spinors in terms of spinorial Lie derivatives that give rise to an anti-symmetric product of vector fields in terms of Lie derivatives on tensor fields. This whole construction is desired for physical reasons related to the statistics obeyed by bosons and fermions. The physical consequences of the interesting algebraic and differential geometric equalities satisfied by 1-form and 2-form Dirac currents will be investigated elsewhere. The main motivation for this stems from the results found in a previous work \cite{Acik Ertem} which relates these objects to physical systems like higher-degree Maxwell fields and Duffin-Kemmer-Petiau fields.

For supergravity theories in different dimensions, the equation satisfied by supergravity Killing spinors gives way to different generalizations of Killing vector fields to higher degree forms. The spinor bilinears of supergravity Killing spinors satisfy the so-called supergravity Killing form equation and it reduces to the Killing equation for the 1-form case. So, in supergravity backgrounds the Killing vector fields have more generalizations to higher degrees other than KY forms. In six and ten dimensional supergravity theories, the solutions of the supergravity Killing form equation have a Lie algebra structure under the Clifford bracket. In five and eleven dimensional cases , the Lie algebra structure is dependent on an extra condition on supergravity Killing forms.

In supergravity backgrounds, a Lie superalgebra that consists of supergravity Killing spinors and Killing vector fields can be constructed and this Killing superalgebra defines an important invariant for the background to obtain a full classification of the supergravity backgrounds \cite{OFarrill HackettJones Moutsopoulos, OFarrill Meessen Philip}. The even part of the Killing superalgebra correspond to the Lie algebra of Killing vector fields and odd part is the space of supergravity Killing spinors. However, as we have seen, the differential forms constructed out of supergravity Killing spinors are not only the duals of Killing vector fields but also other higher degree forms. So, one should include those higher degree forms to obtain a maximal superalgebra of the background. Although there are attempts to construct these maximal superalgebras in general, the exact construction that include the Lie algebra of all higher degree forms has not been achieved yet \cite{OFarrill HackettJones Moutsopoulos Simon, Ertem1, Ertem2}.

As we have indicated in the paper, the spinor bilinears of supergravity Killing spinors which correspond to the higher degree generalizations of Killing vector fields satisfy some Lie algebra structures in different dimensional cases. Those Lie algebra structures can be used to extend the Killing superalgebras and to define the even parts of the maximal superalgebras of the supergravity backgrounds. The construction of the maximal superalgebras by using the Lie algebra structure of the generalizations of Killing vector fields can give hints about the classification problem of supergravity backgrounds.

\begin{acknowledgments}

\"{O}A is grateful to Robin W. Tucker for his guidance, help and kindness during his stay at Lancaster. He also thanks Jonathan Gratus, David Burton and Edward Anderson for insightful discussions on various subjects and owes to Lancaster University for its kind hospitality. \"{U}E thanks Jose M. Figueroa-O`Farrill and Andrea Santi for inspiring discussions on various subjects about supergravity and Lie superalgebras and he also thanks School of Mathematics of The University of Edinburgh for the hospitality and providing a fruitful scientific atmosphere during his stay in Edinburgh. \"{O}A acknowledges partial support from UBED unit of Ankara University. \"{U}E is supported by the Scientific and Technological Research Council of Turkey (T\"{U}B\.{I}TAK) grant B\.{I}DEB 2219.

\end{acknowledgments}

\begin{appendix}

\section{On the Relation between Lie Derivatives and Parallelism}

Although the geometric theory of classical gravity is heavily based on a metric tensor field, the pre-metric notions such as flows and parallelism have their own importance and has effects in both geometrical and physical ways. Since the infinitesimal flows are defined in terms of Lie derivatives of vector fields and infinitesimal parallel transports are given by the covariant derivatives, the tie between them is important in many aspects. In this appendix we underline this relation applied to differential forms and in the presence of a metric structure to Clifford forms and spinors. The primitive notions of exterior differentiation $d$ of forms and Lie differentiation of tensor fields are induced from the differentiable structure of the ambient manifold. An important relation between them that is valid only on form fields is known as the magic formula of \textit{H. Cartan} and is given by
\begin{equation}
{\cal{L}}_{X}=d \circ i _{X}+ i _{X} \circ d
\end{equation}
where $i _{X}$ is the interior derivation with respect to the vector field $X$ and the symbol $\circ$ is the composition of the respective operations which we will omit unless it is necessary. When the notion of a \textit{parallelism} is introduced via a connection $\hat{\nabla}$ then new relations can be built up to define the exterior differentiation in terms of the Lie differentiation. For example one can write $d$ as $e^a\wedge\hat{\nabla}_{X_a}+T^a\wedge i_{X_a}$ for a general connection $\hat{\nabla}$ with torsion $2$-forms $T^a$ or for the special case of a metric-compatible and torsion-free (Levi-Civita) connection $\nabla$ as $e^a\wedge\nabla_{X_a}$ by using any frame basis $\{X_a\}$ and the related coframe basis $\{e^a\}$ with the duality condition $e^a(X_b)=\delta{^a}_{b}$. For the ease of geometrical interpretation, computational convenience and the valid definitions of Lie derivatives on special structures such as Clifford algebras and spinors we give the following data.

If a torsion-free notion of parallelism is defined on a differentiable manifold $M$, then the relation between Lie derivatives and covariant derivatives on differential forms can be put in a form as
\begin{equation}
{\cal{L}}_{X}=\nabla _{X}+e^a \wedge i _{\nabla_{X_a}X} .
\end{equation}
By using the identities $d=e^a\wedge\nabla_{X_a}$ and $[\nabla_{X},i_{Y}]=i_{\nabla_{X}{Y}}$ the result is straightforward. Note that if $X$ is an element of the frame basis, say $X_a$, then one can write ${\cal{L}}_{X_a}=\nabla _{X_a}+{\omega^b}_a \wedge i _{X_b}$. Since ${\cal{L}}_{X}$ and $\nabla_{X}$ behave locally and point-wise respectively with respect to $X$, one concludes that the local character of the former is hidden in the last term.

Let a metric structure $g$, compatible with the torsion-free $\nabla$-parallelism (a Levi-Civita connection) be defined on $M$ and assume that there exists a Killing vector field $K$ i.e. ${\cal{L}}_{K}g=0$ or equivalently $\nabla _{X}\widetilde{K}=\frac{1}{2}i_{X}d\widetilde{K}$ for any $X$. Then, the operation of Lie derivation on Clifford forms $\alpha$ with respect to this Killing vector field can be well defined and is given by
\begin{equation}
{{\cal{L}}_{K}}\alpha={\nabla _{K}}\alpha+\frac{1}{4}[d\widetilde{K},\alpha]_{Cl}
\end{equation}
where $[.,.]_{Cl}$ is the Clifford commutator. Note that since the Clifford algebra is associative, this commutator (namely the Lie commutator defined with respect to the Clifford multiplication) is a derivation of the Clifford algebra. The equality (A3) for the Lie derivation can be proven as follows. By using (A2) and the metric-compatibility condition one can write
\begin{equation}
{\cal{L}}_{K}=\nabla _{K}+e^a \wedge i _{\widetilde{\nabla_{X_a}\widetilde{K}}} \nonumber\\
\end{equation}
and considering the Killing property mentioned above with using two times successively the identity $e^a\wedge i_{X_a}\omega= p \omega$ for any $p$-form $\omega$ we have
\begin{eqnarray}
{\cal{L}}_{K}\alpha&=&\nabla _{K}\alpha+\frac{1}{2}e^a \wedge i _{i_{X_a}d\widetilde{K}}\alpha \nonumber\\
&=&\nabla _{K}\alpha-\frac{1}{2}{i_{X_a}d\widetilde{K}}\wedge i _{X^a}\alpha.
\end{eqnarray}
Then, as a last step, by applying the useful equality $-2i_{X_a}\mu\wedge i_{X^a}\alpha=[\mu,\alpha]_{Cl}$ for a $2$-form $\mu$ and an arbitrary form $\alpha$ we reach the anticipated result (A3) since $d\widetilde{K}$ is a $2$-form.

On the other hand, the definition of the Lie derivative of a spinor field $\psi$ with respect to a Killing vector field $K$ which is given in \cite{Kosmann} is
\begin{equation}
{\pounds_{K}}\psi={\nabla _{K}}\psi+\frac{1}{4}d\widetilde{K}.\psi
\end{equation}
where we separate the spinorial Lie derivative by $\pounds$ from the one ${\cal{L}}$ on forms, but preserve the symbol of covariant differentiation. This definition is a direct result of (A3). The reason is that since the spaces of spinors are (local) left Clifford modules, the left Clifford action part of the Clifford commutator should be kept for the insurance of staying in the spinor space. The usage of Killing vector fields as the arguments of Lie differentiation both on the sections of the Clifford bundle and on the spinor fields is necessary because of the following reasons; 1) The flows of these vector fields preserve the metric tensor field, so that the corresponding Lie derivatives respect the Clifford multiplication, 2) Since the spin structure is determined by the metric tensor field, the conservation of the metric along the Lie transport is mandatory \cite{Dabrowski Percacci}.

\section{Dependence of the Algebra of Dirac Currents on Killing Numbers}

Let $\psi$ and $\phi$ are geometric Killing spinors on $M$ with $V_{\psi}$ and $V_{\phi}$ are corresponding Dirac currents of them. Since the Dirac currents of the geometric Killing spinors are Killing vector fields, we can write the following Lie bracket
\begin{equation}
[V_{\psi}, V_{\phi}]={\cal{L}}_{V_{\psi}}V_{\phi}={\cal{L}}_{V_{\psi}}\widetilde{\widetilde{V_{\phi}}}=\widetilde{{\cal{L}}_{V_{\psi}}\widetilde{V_{\phi}}}.
\end{equation}
By using the degree preserving property of the Lie derivative, one obtains
\begin{eqnarray}
\widetilde{[V_{\psi}, V_{\phi}]}&=&{\cal{L}}_{V_{\psi}}((\phi\bar{\phi})_1)\nonumber\\
&=&({\cal{L}}_{V_{\psi}}(\phi\bar{\phi}))_1\nonumber\\
&=&((\pounds_{V_{\psi}}\phi)\bar{\phi}+\phi(\overline{\pounds_{V_{\psi}}\phi}))_1.
\end{eqnarray}
Geometric Killing spinor equation can be written as
\begin{equation}
\nabla_X\phi=\lambda_{\phi}\widetilde{X}.\phi
\end{equation}
where $\lambda_{\phi}$ is the Killing number dependent on the spinor. In the case of geometric Killing spinors, the definition of the Lie derivative on spinors with respect to a Killing vector corresponding to the Dirac current of another Killing spinor $\psi$ transforms into the following form
\begin{eqnarray}
\pounds_{V_{\psi}}\phi&=&\nabla_{V_{\psi}}\phi+\frac{1}{4}d\widetilde{V_{\psi}}.\phi\nonumber\\
&=&\lambda_{\phi}\widetilde{V_{\psi}}.\phi+\frac{1}{4}d\widetilde{V_{\psi}}.\phi.
\end{eqnarray}
It has been shown in \cite{Acik Ertem} that the exterior derivative of the Dirac current of a geometric Killing spinor is dependent on the chosen inner product and involutions in the Clifford algebra. There are two main classes that give the relations for exterior and co-derivatives of a Dirac current of a Killing spinor in terms of higher or lower degree generalized Dirac currents. Since we only consider 1-form Dirac currents here, the first class contains the following two possibilities; (i) $\lambda$ is real and the involutions $\{j,{\cal{J}}\}$ are chosen as $\{Id, \xi\}, \{Id, \xi^*\}, \{^*, \xi\}$ or $\{^*, \xi^*\}$ and (ii) $\lambda$ is pure imaginary and the involutions are $\{^*, \xi\eta\}$ or $\{^*, \xi\eta^*\}$ (see \cite{Acik Ertem} for the notation). In those cases, we have the following relations
\begin{eqnarray}
d(u\bar{u})_1&=&0\nonumber\\
\delta(u\bar{u})_1&=&-2\lambda_un(u\bar{u})_0
\end{eqnarray}
where $u=\psi$ or $\phi$. The second class contains the following possibilities; (i) $\lambda$ is real and the involutions are $\{Id, \xi\eta\}, \{Id, \xi\eta^*\}, \{^*, \xi\eta\}$ and $\{^*, \xi\eta^*\}$ and (ii) $\lambda$ is pure imaginary and the involutions are $\{^*, \xi\}$ and $\{^*, \xi^*\}$. In those cases, the relations are as follows
\begin{eqnarray}
d(u\bar{u})_1&=&4\lambda_u(u\bar{u})_2\nonumber\\
\delta(u\bar{u})_1&=&0.
\end{eqnarray}
The Dirac currents of the second class correspond to the Killing 1-forms of the background. However, the Dirac currents of the first class do not directly correspond to the Killing 1-forms, but they correspond to the Hodge duals of Killing 1-forms which are called closed conformal Killing-Yano $(n-1)$-forms.
We consider the Dirac currents of second class that correspond to the Killing 1-forms. In that case the Lie derivative of a Killing spinor with respect to a Dirac current can be written in the following form
\begin{equation}
\pounds_{V_{\psi}}\phi=\lambda_{\phi}\widetilde{V_{\psi}}.\phi+\lambda_{\psi}(\psi\bar{\psi})_2.\phi.
\end{equation}
Since the Lie derivative of a Killing spinor is again a Killing spinor, we can compute the 1-form projection of the Clifford product of this new Killing spinor with $\phi$
\begin{equation}
((\pounds_{V_{\psi}}\phi)\bar{\phi})_1=\lambda_{\phi}(\widetilde{X_{\psi}}.(\phi\bar{\phi}))_1+\lambda_{\psi}((\psi\bar{\psi})_2.(\phi\bar{\phi}))_1.
\end{equation}
From the properties of the Clifford product, one finds that
\begin{eqnarray}
((\pounds_{V_{\psi}}\phi)\bar{\phi})_1&=&\lambda_{\phi}\left((\psi\bar{\psi})_1.(\phi\bar{\phi})_0\right)_1+\lambda_{\phi}\left((\psi\bar{\psi})_1.(\phi\bar{\phi})_2\right)_1\nonumber\\
&&+\lambda_{\psi}\left((\psi\bar{\psi})_2.(\phi\bar{\phi})_1\right)_1+\lambda_{\psi}\left((\psi\bar{\psi})_2.(\phi\bar{\phi})_3\right)_1\nonumber\\
&=&\lambda_{\phi}(\phi, \phi)\widetilde{V_{\psi}}+\lambda_{\psi}\left((\psi\bar{\psi})_2.(\phi\bar{\phi})_3\right)_1\nonumber\\
&&+\lambda_{\phi}\left[i_{V_{\psi}}(\phi\bar{\phi})_2-\frac{\lambda_{\psi}}{\lambda_{\phi}}i_{V_{\phi}}(\psi\bar{\psi})_2\right].
\end{eqnarray}
The second term on the right hand side can be written in the following way by using the expansion of the Clifford product in terms of the wedge product
\begin{equation}
\left((\psi\bar{\psi})_2.(\phi\bar{\phi})_3\right)_1=-\frac{1}{2}i_{X_a}i_{X_b}(\psi\bar{\psi})_2\wedge i_{X^a}i_{X^b}(\phi\bar{\phi})_3.
\end{equation}
So we have
\begin{eqnarray}
((\pounds_{V_{\psi}}\phi)\bar{\phi})_1&=&\lambda_{\phi}(\phi, \phi)\widetilde{V_{\psi}}+\lambda_{\psi}\left(i_{X_{I(2)}}(\psi\bar{\psi})_2\wedge i_{X^{I(2)}}(\phi\bar{\phi})_3\right)\nonumber\\
&&+\lambda_{\phi}\left[i_{V_{\psi}}(\phi\bar{\phi})_2-\epsilon_{\psi,\phi}i_{V_{\phi}}(\psi\bar{\psi})_2\right]
\end{eqnarray}
where $\epsilon_{\psi,\phi}=\frac{\lambda_{\psi}}{\lambda_{\phi}}$. The spinor dual of the Lie derivative of a Killing spinor with respect to a Dirac current is
\begin{equation}
\overline{\pounds_{V_{\psi}}\phi}=\lambda_{\phi}^j\bar{\phi}.\widetilde{V_{\psi}}^{\cal{J}}+\lambda_{\psi}^j\bar{\phi}.[(\psi\bar{\psi})_2]^{\cal{J}},
\end{equation}
hence we have
\begin{equation}
\phi\overline{\pounds_{V_{\psi}}\phi}=\lambda_{\phi}^j(\phi\bar{\phi}).\widetilde{V_{\psi}}^{\cal{J}}+\lambda_{\psi}^j(\phi\bar{\phi}).[(\psi\bar{\psi})_2]^{\cal{J}}
\end{equation}
and we can write the following 1-form projection
\begin{eqnarray}
(\phi\overline{\pounds_{V_{\psi}}\phi})_1&=&\lambda_{\phi}^j(\phi, \phi)\widetilde{V_{\psi}}^{\cal{J}}+\frac{\lambda_{\psi}^j}{2}i_{X_{I(2)}}[(\psi\bar{\psi})_2]^{\cal{J}}i_{X^{I(2)}}(\phi\bar{\phi})_3\nonumber\\
&&-\lambda_{\phi}^j\left[i_{\widetilde{V_{\psi}}^{\cal{J}}}(\phi\bar{\phi})_2-\epsilon_{\psi,\phi}i_{V_{\phi}}[(\psi\bar{\psi})_2]^{\cal{J}}\right].
\end{eqnarray}
Now we can write the metric dual of the Lie bracket of two Dirac currents of Killing spinors from (B2), (B9) and (B14)
\begin{eqnarray}
\widetilde{[V_{\psi}, V_{\phi}]}&=&(\phi, \phi)\left(\lambda_{\phi}\widetilde{V_{\psi}}+\lambda_{\phi}^j\widetilde{V_{\psi}}^{\cal{J}}\right)\nonumber\\
&&-\frac{1}{2}\left[\left(\lambda_{\psi}i_{X_{I(2)}}(\psi\bar{\psi})_2+\lambda_{\psi}^ji_{X_{I(2)}}[(\psi\bar{\psi})_2]^{\cal{J}}\right)i_{X^{I(2)}}(\phi\bar{\phi})_3\right]\nonumber\\
&&+\lambda_{\phi}\left[i_{V_{\psi}}(\phi\bar{\phi})_2-\epsilon_{\psi,\phi}i_{V_{\phi}}(\psi\bar{\psi})_2\right]-\lambda_{\phi}^j\left[i_{\widetilde{V_{\psi}}^{\cal{J}}}(\phi\bar{\phi})_2-\epsilon_{\psi,\phi}i_{V_{\phi}}[(\psi\bar{\psi})_2]^{\cal{J}}\right].
\end{eqnarray}
If we choose $\lambda$ real and ${\cal{J}=\xi\eta}$, this reduces to
\begin{eqnarray}
\widetilde{[V_{\psi}, V_{\phi}]}=2\lambda_{\phi}\left[i_{V_{\psi}}(\phi\bar{\phi})_2-\epsilon_{\psi, \phi}i_{V_{\phi}}(\psi\bar{\psi})_2\right]
\end{eqnarray}
and since $\lambda$ is real, the choice of $j$ does not effect the results. So, we have to choose $\epsilon_{\psi, \phi}=1$ because of the antisymmetry property of left hand side. This means that we must have $\lambda_{\psi}=\lambda_{\phi}=a$ or $\lambda_{\psi}=\lambda_{\phi}=-a$ which gives topologically distinct simultaneous subalgebras of the symmetry algebra. Hence, we obtain the following equality
\begin{eqnarray}
\widetilde{[V_{\psi}, V_{\phi}]}&=&2\lambda\left[i_{V_{\psi}}(\phi\bar{\phi})_2-i_{V_{\phi}}(\psi\bar{\psi})_2\right]\nonumber\\
&=&2\lambda\left(i_{V_{\psi}}(\phi\bar{\phi})-i_{V_{\phi}}(\psi\bar{\psi})\right)_1.
\end{eqnarray}
By using the expansion of the Clifford product in terms of wedge product and interior derivative;
\begin{eqnarray}
\widetilde{V_{\psi}}.(\phi\bar{\phi})&=&\widetilde{V_{\psi}}\wedge\eta (\phi\bar{\phi})+i_{V_{\psi}}(\phi\bar{\phi})\nonumber\\
(\phi\bar{\phi}).\widetilde{V_{\psi}}&=&\widetilde{V_{\psi}}\wedge\eta (\phi\bar{\phi})-i_{V_{\psi}}(\phi\bar{\phi}),
\end{eqnarray}
one can show that (B17) can be written as follows
\begin{equation}
\widetilde{[V_{\psi}, V_{\phi}]}=\left(\widetilde{V_{\psi}}.(\phi\bar{\phi})-\widetilde{V_{\phi}}.(\psi\bar{\psi})\right)_1.
\end{equation}
From the Killing spinor equation, one can also find that
\begin{equation}
\left(\nabla_{V_{\psi}}(\phi\bar{\phi})\right)_1=\lambda\left(\widetilde{V_{\psi}}.(\phi\bar{\phi})\right)_1+\lambda\left((\phi\bar{\phi})^{\eta}\widetilde{V_{\psi}}^{\eta}\right)_1.
\end{equation}
So ,we have the following identity for the Lie bracket of two Dirac currents of Killing spinors
\begin{equation}
\widetilde{[V_{\psi}, V_{\phi}]}=\left(\nabla_{V_{\psi}}(\phi\bar{\phi})-\nabla_{V_{\phi}}(\psi\bar{\psi})\right)_1
\end{equation}
and this corresponds to the definition of the Lie bracket in terms of the Lie derivative. Hence, to obtain the consistency of the Lie bracket of Dirac currents of geometric Killing spinors, we have to use the fact that the Killing numbers of two spinors must be equal to each other. This result has been seen from the algebraic computation of the Lie bracket of Dirac currents which can not be seen from the geometric computation given in section 2A.

\section{Clifford Algebraic Relations for Dirac Currents and Some Important Identities}

The geometric relation given in terms of Lie differentiation between the $1$-form Dirac currents of two geometric Killing spinors can be cast into an algebraic relation in terms of the Clifford commutator of the corresponding $1$ and $2$-form Dirac currents. Before and after reaching this result, we will deduce some mixed (both algebraic and geometric) identities satisfied by these specific Dirac currents.

i) By considering (18)
\begin{equation}
\mathcal{L}_{V_{\psi}}\widetilde{V_{\phi}}=\widetilde{[V_{\psi},V_{\phi}]}
\end{equation} 
and manipulating the left hand side by using (A3) with $K=V_{\psi}$ and also using the identity $d\widetilde{V_{\psi}}=4\lambda (\psi \overline{\psi})_{2}$
one can reach
\begin{equation}
\mathcal{L}_{V_{\psi}}\widetilde{V_{\phi}}=\nabla_{V_{\psi}}\widetilde{V_{\phi}}+\lambda[(\psi \overline{\psi})_{2},\widetilde{V_{\phi}} ]_{Cl}
\end{equation}
since the Clifford commutator is anti-symmetric. Then, considering the vanishing torsion condition and the metric compatibility for the right hand side of (C1) and equating it to the right hand side of (C2) we deduce
\begin{equation}
\nabla_{V_{\phi}}\widetilde{V_{\psi}}=\lambda[\widetilde{V_{\phi}},(\psi \overline{\psi})_{2}]_{Cl}.
\end{equation}
Additionally, this can be recast into the following form by using the definition of Clifford multiplication
\begin{equation}
\nabla_{V_{\phi}}\widetilde{V_{\psi}}=2\lambda i_{V_{\phi}}(\psi \overline{\psi})_{2}.
\end{equation}
If the Dirac currents satisfy a Lie algebra, namely $\widetilde{[V_{\psi},V_{\phi}]}=C_{\psi,\phi,\zeta}\widetilde{V_{\zeta}}$, then by employing the last equality simultaneously with the torsion-free property of $\nabla$, the following algebraic equality can be obtained
\begin{equation}
2\lambda (i_{V_{\psi}}(\phi \overline{\phi})_{2}-i_{V_{\phi}}(\psi \overline{\psi})_{2})=C_{\psi,\phi,\zeta}\widetilde{V_{\zeta}}
\end{equation}
or in terms of Clifford commutators
\begin{equation}
\widetilde{V_{\zeta}}=\frac{\lambda}{C_{\psi,\phi,\zeta}}\left([\widetilde{V_{\psi}},(\phi \overline{\phi})_{2}]_{Cl}-[\widetilde{V_{\phi}},(\psi \overline{\psi})_{2}]_{Cl}\right).
\end{equation}
This is the 1-form Dirac current counterpart of (30). In fact the combination of (30) and (C6) gives the following equality
\begin{equation}
\widetilde{V_{\zeta}}+(\zeta\bar{\zeta})_2=\frac{\lambda}{C_{\psi, \phi, \zeta}}\left[\widetilde{V_{\psi}}+(\psi\bar{\psi})_2, \widetilde{V_{\phi}}+(\phi\bar{\phi})_2\right]_{Cl}.
\end{equation}

ii) If one takes the exterior derivative of both sides of (C4) and uses (B6) 
\begin{equation}
d\nabla_{V_{\phi}}\widetilde{V_{\psi}}=\frac{1}{2}d i_{V_{\phi}}d\widetilde{V_{\psi}}
\end{equation}  
and applying the Cartan formula, the property $d\circ {\cal{L}}_{X}= {\cal{L}}_{X}\circ d$ and (C1), then the result is
\begin{equation}
2 d\nabla_{V_{\phi}}\widetilde{V_{\psi}}=d{\cal{L}}_{V_{\phi}}\widetilde{V_{\psi}}=d\widetilde{[V_{\phi}, V_{\psi}]}.
\end{equation} 
Since the right hand side of this equation is anti-symmetric with respect to $\psi\leftrightarrow \phi$ this implies that the $\psi\leftrightarrow \phi$-symmetric part of left hand side
must vanish and that gives rise to the following conservation laws
\begin{equation}
d(\nabla_{V_{\phi}}\widetilde{V_{\psi}}+\nabla_{V_{\psi}}\widetilde{V_{\phi}})=0
\end{equation} 
which also from (B4) implies
\begin{equation}
d(i_{V_{\phi}}(\psi \overline{\psi})_{2}+i_{V_{\psi}}(\phi \overline{\phi})_{2})={\cal{L}}_{V_{\phi}}(\psi \overline{\psi})_{2}+{\cal{L}}_{V_{\psi}}(\phi \overline{\phi})_{2}=0.
\end{equation}
For the special case of $\phi=\psi$
\begin{equation}
d(\nabla_{V_{u}}\widetilde{V_{u}})=0
\end{equation} 
where $u \in \{\phi,\psi\}$, this implies
\begin{equation}
d(i_{V_{u}}(u \overline{u})_{2})={\cal{L}}_{V_{u}}(u \overline{u})_{2}=\left(({\pounds}_{V_{u}}u) \overline{u}+u (\overline{{\pounds}_{V_{u}}u})\right)_2=0.
\end{equation}

iii) Another property that worths mentioning is that, since $(\kappa \overline{\kappa})_{2}$ is a closed conformal Killing-Yano $2$-form for $\kappa \in \{\phi, \psi\}$ we can write $\nabla_{V_{\phi}}(\psi \overline{\psi})_{2}=-\frac{1}{n-1}\widetilde{V_{\phi}} \wedge \delta (\psi \overline{\psi})_{2}$ and from the equality $\delta (\psi \overline{\psi})_{2}=-2 \lambda (n-1)\widetilde{V_{\psi}}$ we find 
\begin{equation}
\nabla_{V_{\phi}}(\psi \overline{\psi})_{2}=2 \lambda \widetilde{V_{\phi}} \wedge \widetilde{V_{\psi}},
\end{equation}
which is a kind of restricted Sasaki equation that is used in the definition of Sasakian manifolds. In particular, for $\phi=\psi$
\begin{equation}
\nabla_{V_{\psi}}(\psi \overline{\psi})_{2}=0.
\end{equation}

iv) A relation similar to (30) can be obtained by using the previous results. The first equality of (C8) with the orders of Lie derivative and exterior derivative are changed and (B6) for Dirac currents results to the identity $2 \lambda {\cal{L}}_{V_{\phi}}(\psi \overline{\psi})_{2}=d \nabla_{V_{\phi}}\widetilde{V_{\psi}}$. Rewriting this with $\phi\leftrightarrow\psi$ and subtracting from the first one, it reads 
\begin{equation}
2 \lambda \left({\cal{L}}_{V_{\phi}}(\psi \overline{\psi})_{2}-{\cal{L}}_{V_{\psi}}(\phi \overline{\phi})_{2}\right)=d \widetilde{[V_{\phi}, V_{\psi}]}.
\end{equation}
Using (A3), (B6), (C15) and the definition of Clifford multiplication one gets
\begin{equation}
4 \lambda^2 \left([\widetilde{V_{\phi}}, \widetilde{V_{\psi}} ]_{Cl}+[(\phi \overline{\phi})_{2},(\psi \overline{\psi})_{2}]_{Cl}\right)
=d \widetilde{[V_{\phi}, V_{\psi}]}
\end{equation}
which is the key equation for the algebraic closure condition (30). Moreover, by using (C11) the combination of the last two equations gives
\begin{equation}
{\cal{L}}_{V_{\phi}}(\psi \overline{\psi})_{2}=\lambda\left([\widetilde{V_{\phi}}, \widetilde{V_{\psi}} ]_{Cl}+[(\phi \overline{\phi})_{2},(\psi \overline{\psi})_{2}]_{Cl}\right).
\end{equation}

 \end{appendix}

%\references%

\end{document}